# Application of Machine Learning to Mechanical Properties of Copper Graphene Composites

Milan Rohatgi[1], Amir Kordijazi[2*]
[1]Henry M. Gunn High School, Palo Alto, CA 94306, USA
[2]Colleges of Nanoscale Science and Engineering, SUNY Polytechnic Institute, Albany, NY 12203, USA

## Abstract

While copper-graphene (Cu/Gr) composites have been promising materials due to their theoretically high strength and conductivity, their design has been hampered by the large number of variables affecting their properties. We applied four different Machine Learning (ML) models to manually collected datasets compiling the yield strength and ultimate tensile strength of graphene-reinforced copper composites processed with powder metallurgy techniques. Our results indicate that ML models can predict the mechanical properties of Cu/Gr composites with satisfactory accuracy. Feature analysis provided new insights into the most important factors that affect these properties.

**KEYWORDS:** Copper, Graphene, Predictive modeling, Machine learning

## 1 Introduction

Composite materials have the potential to solve many problems in materials science. However, their efficient design is complicated by the impact of an enormous number of variables: the identity of the material used, the geometrical arrangement of the components, the manufacturing process, and much more. Data-driven approaches like machine learning (ML) are powerful methods to derive useful predictive information in design challenges that deal with a large number of input variables and noisy or sparse datasets (such as published experimental datasets on Cu/Gr composites). ML approaches were first applied to the discovery of chemical compounds with new or bespoke properties [1–3] and, more recently, to the design of composite materials [4,5]. For example, ML approaches have been applied to predict wetting properties of iron-based and aluminum-based composites [6,7], tribological properties of aluminum-graphite composites [8–10], mechanical properties of carbon fiber reinforced composites [11], and the compressive strength of concrete [12].

There has been increasing interest in the incorporation of graphene as a reinforcement in copper due to its exceptionally high modulus, strength, electrical conductivity, and thermal conductivity. The addition of graphene can also reduce the weight of the composite due to the much lower density of graphene compared to copper [13]. However, despite the promise of copper-graphene (Cu/Gr) composites, they have not found significant applications in engineering, with experimentally measured properties often falling short of theoretical predictions [13]. In addition, myriad variables seem to influence the properties of Cu/Gr composites, making it time-consuming to use iterative trial-and-error-based approaches and computationally prohibitive to use physics-based simulations or calculations to exhaustively search the design space [4].

We used a variety of ML models to predict the yield strength and ultimate tensile strength of Cu/Gr composites. Yield strength and ultimate tensile strength are two commonly reported metrics to measure a material's mechanical strength under stress. Yield strength (in Pascals or Pa) describes the amount of stress at which a material begins to undergo measurable plastic deformation (as opposed to elastic deformation), and ultimate tensile strength (in Pa) is the maximum amount of stress a material can experience before breaking. A variety of variables affect these two properties, making empirical or intuition-based design strategies challenging. Because the manufacturing process can have a large effect on the mechanical strength of Cu/Gr composites, we chose to only analyze composites made using powder metallurgy. We sought to optimize multiple models and then compare their respective performances to find the best ML model for predicting either yield strength or ultimate tensile strength. Finally, feature importance analysis was used to rank the compositional and processing variables in order of predictive importance.

## 2 Materials and Methods

### 2.1 Data Collection and Model Variables

Data was collected from multiple papers in the literature given the importance of both quantity and diversity of data in training models [14]. Using forty publications, 145 points were collected for yield strength and 88 points for tensile strength [13,15–51]. Single-source data may produce models that converge on a pattern that is not representative of Cu/Gr as a whole and would thus be irrelevant for data collected in a different setting. Three types of input variables were used: graphene content, type of graphene, and processing route (**Table 1**). However, processing route was split into multiple features as many composites were created using more than one. All processing route parameters were categorical (yes/no) with the exception of ball milling speed/time in yield strength data, which was numerical. Further, type of graphene (list types) was categorical, while graphene content was collected numerically.

**Table 1: Input and output variables for yield strength and tensile strength data**

| Input Variables | Output Variables |
|---|---|
| vol% graphene, type of graphene, ball mill time, ball mill speed, sonication, sintering, hot pressing, stirring, molecular level mixing, hot pressing, in-situ growth, heat treatment, rolling, electrostatic self-assembly | Yield Strength |
| vol% graphene, type of graphene, ball milling, sonication, spark plasma sintering, hot pressing, in-situ growth, stirring, electrostatic self-assembly, mixing, hot rolling | Tensile Strength |

**2.2 Machine Learning Models**

We developed four ML regression models in Python (using the Sci-kit learn library) [52] to predict values for yield strength and tensile strength for Cu/Gr composites: Random Forest (RF), Artificial Neural Network (ANN), Gradient Boosting Machine (GBM), and K-Nearest Neighbor (KNN). Different models were used for predicting tensile strength and yield strength to allow for parameter tuning specific to each prediction.

Random Forests are an ensemble model that combines the predictions of many decision trees to make a final prediction. Each decision tree makes its prediction by splitting the data up at nodes based on certain features. For example, one node may split the data based on the type of graphene used. This happens multiple times at different nodes until the model is able to make a unique prediction based on all input parameters. The results of all of the decision trees are then averaged to form a final predicted value. This method protects against inconsistencies or outliers in the performance of a single decision tree. In addition, it is comparatively fast compared to other ML models, so works well with large data sets. Important features to optimize in an RF are the number of features and the number of decision trees.

Artificial Neural Networks attempt to mimic the decision patterns of the human brain. Prediction happens across multiple layers. The first layer is the input layer, where all of the parameters are introduced as individual nodes. The data is then put through multiple hidden layers, each with multiple nodes. The values at each of these nodes are determined by a combination of every node in the previous layer, but can be altered by weights and biases so that each node has a different value. After going through each hidden layer, the final prediction is shown in the output layer. As more and more data goes through the ANN, the weights and biases between each layer are adjusted to make better predictions each time. The particular type of ANN used for our studies was a Multilayer perceptron (MLP). This type of model performs well with regression problems that have nonlinear trends in data. Important features to optimize in an ANN are the activation function, regularization parameter, number of hidden layers, and number of nodes per hidden layer.

Gradient Boosting Machines are also ensemble methods that utilize decision trees. However, the model is different from a Random Forest because each successive tree makes a decision based on the performance of the previously created trees. Further, each tree uses a different set of features to split the data at nodes. After each tree has made its predictions, they are averaged out to get a final predicted value. Important features to optimize a GBM are the learning rate and the number of boosting stages (n_estimators).

K-Nearest-Neighbor Regressors make a prediction based on the closest data points in the training set. These distances are determined multidimensionally, with each dimension representing a different parameter. The number of nearest data points considered is determined by the developer and is crucial to the performance of the KNN. Further, weights can be assigned based on how close each considered point is to the test point. The final value in a regression KNN is determined by averaging the values of the nearest points, including weights if applicable. Important features to optimize in a KNN are the number of neighbors and weights.

**2.3 Data Preprocessing and Parameter Tuning**

Multiple steps were taken to ensure that the data was in optimal condition before being used for ML. Scikit-learn's OneHotEncoder was used on categorical parameters so that they were in a functional format for the ML models. Next, the "Standard Scaler" was used to normalize data and scale features. The data was also split into two sets; one to train the models, and one to test their performance. 75% (109 points for yield strength, 66 points for UTS) of the data went into the training set, while the other 25% was used for testing. Parameter tuning was done using GridSearch from Scikit-learn. The best parameters were determined by inputting a series of parameter combinations through multiple cross-validations (divisions of training and test sets). Based on the average performance in these cross-validations, the parameters in **Tables 2 and 3** were used.

**Table 2: Parameters for ML models (Yield Strength)**

| | |
|---|---|
| RF | max_features = 3, n_estimators = 500 |
| KNN | n_neighbors = 2, weights = 'uniform' |
| GBM | learning_rate = .5, loss = 'absolute_error', n_estimators = 200 |
| ANN | activation = 'relu', hidden_layer_sizes = (15,15,15), solver = 'adam', alpha = .0001 |

**Table 3: Parameters for ML models (Ultimate Tensile Strength)**

| RF | max_features = 3, n_estimators = 150 |
|---|---|
| KNN | n_neighbors = 2, weights = 'uniform' |
| GBM | learning_rate = .1, loss = 'absolute error', n_estimators = 100 |
| ANN | activation = 'relu', hidden_layer_sizes = (25,25,25), solver = 'adam', alpha = .001 |

## 3 Results and Discussion

**Table 4: Statistical Metrics for ML Model Performance on Yield Strength**

| Yield Strength | $R^2$ | MSE | MAE |
|---|---|---|---|
| ANN | .8366 | .1634 | .2809 |
| GBM | .9073 | .0927 | .2103 |
| KNN | .8571 | .1429 | .2423 |
| RF | .9214 | .0786 | .1945 |

**Table 5: Statistical Metrics for ML Model Performance on UTS**

| UTS | $R^2$ | MSE | MAE |
|---|---|---|---|
| ANN | .8948 | .1051 | .1990 |
| GBM | .8932 | .1068 | .2124 |
| KNN | .8238 | .1762 | .2755 |
| RF | .8961 | .1058 | .1860 |

Three different metrics were used to evaluate the performance of the machine learning models. $R^2$ score, a value between 0 and 1, describes how well a regression model fits a set of data; in general, a higher $R^2$ value indicates better performance. Mean Squared Error (MSE) and Mean Absolute Error (MAE) both measure the average distance between the predicted and actual data. However, as the name suggests, MSE squares this distance, which punishes models for larger errors in prediction. Small values for both MSE and MAE indicate good model performance.

Based on Tables 1 and 2, all ML models performed satisfactorily in prediction, with high $R^2$ values and low MSE and MAE values. In Yield Strength predictions, Random Forest (RF) performed significantly better than the other three models, with the highest $R^2$ value and the lowest MSE and MAE values. In UTS predictions, all models except the K-Nearest-Neighbor (KNN) performed similarly; still, the Random Forest model performed the best, with the highest $R^2$ value and lowest MAE value.

Visual representations of the performance of Random Forest (RF) on both Yield Strength and UTS are shown in Figures 1-4.

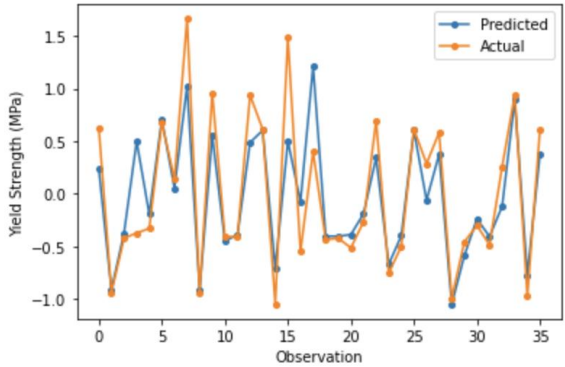

Figure 1: Predicted vs. Actual Values for Yield Strength

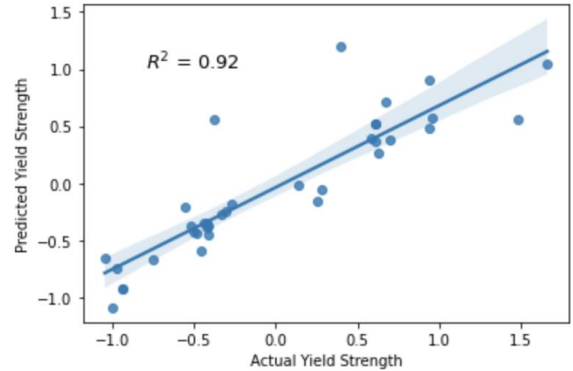

Figure 2: Confidence Interval for Yield Strength Predictions

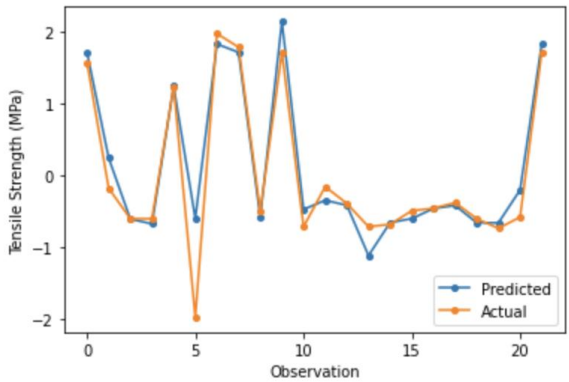

Figure 3: Predicted vs. Actual Values for Yield Strength

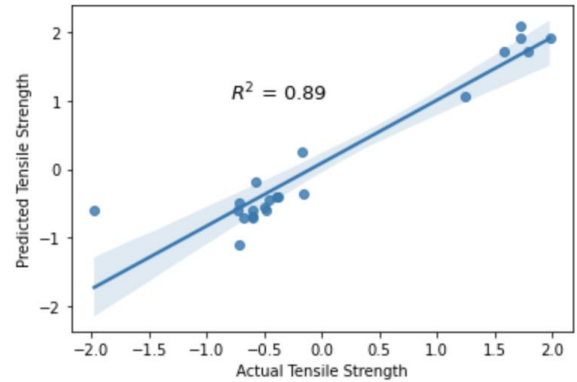

Figure 4: Predicted vs. Actual Values for Yield Strength

Additional analysis was also conducted to find the most important features for predictions of both properties. Note that the differences in number of features is due to some processing routes only present in literature reporting values of one property.

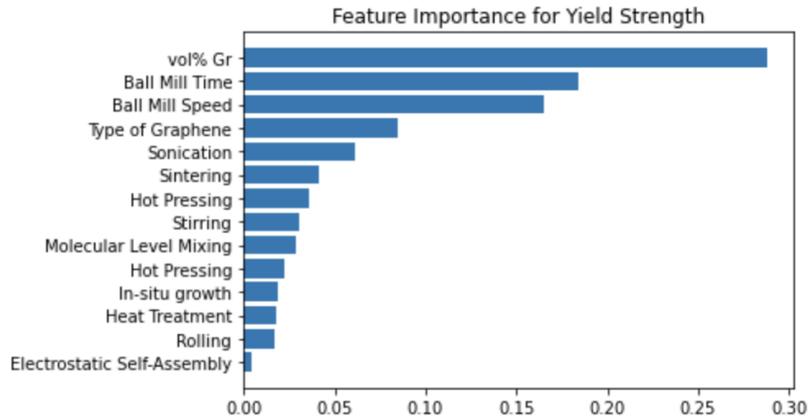

Figure 5: Feature Importance Analysis for Yield Strength

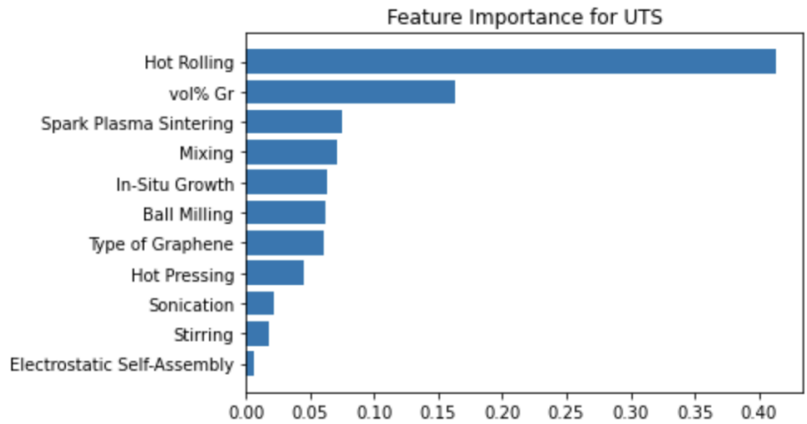

Figure 6: Feature Importance Analysis for UTS

Feature analysis shows that for yield strength, vol% graphene was the most important factor, followed by ball milling speed and time. For UTS, hot rolling was the most important, followed by vol% graphene. Overall, it appears that vol% graphene has a big impact on the mechanical properties of Cu/Gr composites, since it ranks high for both yield strength and UTS. It is important to note that ball milling was entered as a single categorical feature in the UTS data because speed and time were not always reported in the literature. Thus, it is possible that ball milling would have an increased effect on UTS if speed and time were specified.

Theoretical composite mechanics predict that mechanical properties increase with volume percent reinforcement (in this case, graphene). It is possible to calculate the increase in selected properties of reinforced composites, including the modulus, if the reinforcement is aligned and is in the shape of a fiber. However, graphene is flake-shaped and does not align in a copper matrix, making it difficult to calculate even simple mechanical properties. Second, there are no simple theoretical formulations to calculate the effects of processing variables such as ball milling, mixing, and sintering. If the structural changes due to processing are quantitatively characterized, then some changes in mechanical properties may be theoretically estimated. However, calculating the combined effects of changes in composition and processing route is nearly impossible.

Given these challenges, the results shown by feature analysis provide novel insight into copper-graphene composites. While the importance of vol% graphene might be expected, the effects of ball milling would not have been estimated from theory. Likewise, while hot rolling is known to increase mechanical properties due to changes in the microstructure of copper matrices and the alignment of graphene, theoretical considerations would not have predicted it to be the most important variable in predicting UTS.

**4 Conclusion:**
The ML approach provides useful (and unexpected) guidance for optimization of the mechanical properties of copper graphene composites through changes in composition and processing. More generally, our analysis supports the notion that ML approaches can be invaluable for predicting the properties of composite materials, which are often a complex combination of multiple processing and compositional variables.

**Conflict of interest:** All authors declare no conflict of interest.
**Funding Information and Acknowledgement:** N/A
**Data Availability Statement:** Data will be provided by the corresponding author upon request.